# Mathematical Theory of Locally Coherent Quantum Many-Body Fermionic System

Xindong Wang[1]

Sophyics Technology, LLC

## Abstract

Recently Wang and Cheng proposed a self-consistent effective Hamiltonian theory (SCEHT) for many-body fermionic systems (Wang & Cheng, 2019). This paper attempts to provide a mathematical foundation to the formulation of the SCEHT that enables further study of excited states of the system in a more systematic and theoretical manner. Gauge fields are introduced and correct total energy functional in relations to the coupling gauge field is given. We also provides a Monte-Carlo numerical scheme for the search of the ground state that goes beyond the SCEHT.

## Introduction

Recently Wang and Cheng proposed a SCEHT (Wang & Cheng, 2019) for many-body fermionic systems that is based on a separable variational wavefunction between a local system and its environment and subsequently applies a symmetry-inspired self-consistent ansatz for the final solution of the ground state. The unusual consequence of the theory is the so-called single fermion coherent condensate exists within the framework of the theory. In this paper, we provide a more rigorous mathematical foundation for SCEHT that will help to address the skepticism on the more intuitive variational approach, and more importantly, to provide a unified theory for many-body physics and local gauge field theory (Yang & Miles, 1954).

We will first introduce fermion coherent state representation for the ground state of SCEHT. The gauge degree of freedom for each local subspace is then incorporated as a local gauge field to the theory. Next, we present a total energy functional that includes both the on-site and inter-site couplings, the presentation of the latter is in terms of the gauge field introduced. Finally, we presented a variational numerical algorithm for the search of the ground state that goes beyond the SCEHT.

## Fermion Coherent State Representation of Many-Body Fermion State

If a local Fock space is defined by $n$ pairs of fermion creation and annihilation operators ($n=2$ for single-band Hubbard model, for example), we can use the Grassmann number (Negele & Orland, 1987) expansion as the representation of any local many-body states. This expanded mathematical complexity allows us to represent the full system state using the following site permutation symmetric form which the original variational wavefunction in (Wang & Cheng, 2019) did not explicitly display:

---

[1] Email: xdwang98@yahoo.com

$$|\Psi(\xi)\rangle = \prod_x f_x(\{\hat{\psi}_\alpha^\dagger(x)\xi_\alpha(x)\})|0\rangle \qquad [1.1]$$

It is noted that the above coherent state representation of the full system state has the following site permutation invariance:

$$f_x(\{\hat{\psi}_\alpha^\dagger(x)\xi_\alpha(x)\})f_{x'}(\{\hat{\psi}_\alpha^\dagger(x')\xi_\alpha(x')\}) = f_{x'}(\{\hat{\psi}_\alpha^\dagger(x')\xi_\alpha(x')\})f_x(\{\hat{\psi}_\alpha^\dagger(x)\xi_\alpha(x)\}) \qquad [1.2]$$

The inner product of two states in Fock space can be calculated as

$$\langle\Psi'|\Psi\rangle = \int \prod_{x,\alpha} d\xi_\alpha^*(x)d\xi_\alpha(x) \exp\left(-\sum_{x,\alpha}\xi_\alpha^*(x)\xi_\alpha(x)\right)\langle\Psi'(\xi)|\Psi(\xi)\rangle \qquad [1.3]$$

And similarly for operators, we have

$$\langle\Psi'|\hat{O}|\Psi\rangle = \int \prod_{x,\alpha} d\xi_\alpha^*(x)d\xi_\alpha(x) \exp\left(-\sum_{x,\alpha}\xi_\alpha^*(x)\xi_\alpha(x)\right)\langle\Psi'(\xi)|\hat{O}|\Psi(\xi)\rangle \qquad [1.4]$$

The expectation value of $\hat{O}$ (a c complex number in general) under a given state is then

$$\frac{\langle\Psi|\hat{O}|\Psi\rangle}{\langle\Psi|\Psi\rangle}$$

The above coherent site permutable representation can be used to be basis function for the generalized Fock space and any state can be expressed as

$$|\Psi(\xi)\rangle = \prod_x f_x\left((\{\hat{\psi}_\alpha^\dagger(x)\xi_\alpha(x)\})\right)|0\rangle = \prod_x \sum_m c_{nm}(x)\Psi_m(\{\hat{\psi}_\alpha^\dagger(x)\xi_\alpha(x)\})|0\rangle, \qquad [1.5]$$

where $\{\Psi_n(\{\hat{\psi}_\alpha^\dagger(x)\xi_\alpha(x)\})\}$ form a set of localized many-body states that form a complete basis for the local subspace.

In (Wang & Cheng, 2019) the periodic symmetry (translational symmetry in continuum limit) ansatz is equivalent in current formulation to

$$f_x = e^{i\theta(x)}f, \quad \text{where } \theta(x) \text{ is a scalar.} \qquad [1.6]$$

### Local Gauge Fields and Elementary Excitations

In this section, we will demonstrate that equation [1.5] is equivalent to a local (non-abelian) gauge field formulation of the full many-body fermion system.

First, we will show that for any pure state of the full system, the reduced local density matrix has some important properties.

Start with a pure state of the whole system $|\Psi\rangle$, the density matrix of the whole system is defined as

$$\hat{\rho} = |\Psi\rangle\langle\Psi| \qquad [2.1]$$

Partitioning the whole system into a subsystem S and remaining environment E, we have the following expansion for $|\Psi\rangle$

$$|\Psi\rangle = \sum_{n,\mu} c_{n,\mu} |S,n\rangle|E,\mu\rangle, \qquad [2.2]$$

where $\{|S,n\rangle\}$ and $\{|E,\mu\rangle\}$ are orthonormal basis for the subsystems S and E.

The reduce local density matrix is then defined as

$$\hat{\rho}_S = \mathcal{T}r_\mathcal{E}(|\Psi\rangle\langle\Psi|) = \sum_{nm}\left(\sum_\mu c_{n,\mu} c^*_{m,\mu}\right)|S,n\rangle\langle S,m| \qquad [2.3]$$

Note that the operator $\hat{\rho}_S$ is Hermitian and its eigenvalues are non-negative and summed to 1:

$$\mathcal{T}r_S(\hat{\rho}_S) = 1 \qquad [2.4]$$

Given the reduced density matrix, any local operator expectation value can be calculated as

$$\langle \hat{O}_S \rangle = \mathcal{T}r_S(\hat{O}_S \hat{\rho}_S) = \mathcal{T}r_S(\hat{O}_S \hat{\rho}_S) \qquad [2.5]$$

The eigenvalue spectrum of the reduced density matrix is invariant under local unitary transformations:

$$\hat{U}(x) = e^{i\mathcal{A}(x)}, \quad \text{where } \mathcal{A}(x) \text{ is Hermitian.} \qquad [2.6]$$

Next, we will show that if the pure state in [2.1] is an eigenstate of the whole system Hamiltonian, the reduced effective Hamiltonian is related to the reduced density matrix in a proportional manner.

The effective Hamiltonian for subsystem $S$ is

$$\mathcal{H}^S_{eff}(\Psi) = \mathcal{T}r_\mathcal{E}\{\hat{\rho}\mathcal{H}\} = \mathcal{E}_\Psi \mathcal{T}r_\mathcal{E}\{\hat{\rho}\} = \mathcal{E}_\Psi \rho_S \qquad [2.7]$$

Note that if $|\Psi\rangle = |\mathcal{G}\rangle$ is the ground state of the whole fermion system with non-zero fermion particle number and/or density, the eigenvalue has to be negative (i.e. $\mathcal{E}_g < 0$). This is because we have

$$\mathcal{H} = \hat{H} - \mu \mathcal{N}$$

And $\hat{H}$ is normal ordered operator, hence

$$\mathcal{H}|0\rangle = 0, \quad \mathcal{N}|0\rangle = 0, \quad \mathcal{H}|0\rangle = 0$$

Therefore, the ground state with non-zero expectation value of particle number will have to have lower energy than the vacuum state.

Another important observation is that the ground state of $\mathcal{H}_{eff}^S(\mathcal{G})$ is also the most probable state of $\rho_S$.

Note that if the effective Hamiltonian has a non-degenerate ground state, the local density matrix constructed from

$$|\mathcal{G}_{eff}\rangle\langle\mathcal{G}_{eff}|$$

is automatically ensured to have the most probable state of probability 1 and the self-consistency condition is preserved.

Next, we present the two Ansatz that will form the foundation for a gauge field theory for both ground state and excited states of the whole system.

Locality Ansatz:

*Ansatz of Locality: Any physical eigenstate of the whole system corresponds to a unique $\hat{U}(x)$, up to a global unitary transformation, where the tensor product of the subsystems identified by the set of $\{x\}$ defines the whole system which is assumed to be connected topologically.*

And the Continuity Ansatz:

*Ansatz of Continuity: Any excited state of the whole system corresponds to a $i\mathcal{A}_r(x) = \log\left(\hat{U}(x)\hat{U}_0^\dagger(x)\right)$, that is continous in $x$.*

$\hat{U}_0^\dagger(x)$ corresponds to the ground state that may or may not be continuous.

Using these two principles, we can cast solving the pure many body fermion system to solving the coupled problem between local system and the local gauge field (defined as $i\mathcal{A}_r(x) = \log\left(\hat{U}(x)\hat{U}_0^\dagger(x)\right)$) associated with each local system.

Here the difference between the $\mathcal{A}(x)$ in Eq.[2.6] and the $\mathcal{A}_r(x)$ is that the latter is the so-called renormalized gauge field.

With this transformation, the total Hamiltonian becomes a functional of the gauge fields:

$$\widehat{\mathcal{H}} = \sum_x \widehat{U}(x)\left(\widehat{H}(x) - \mu\hat{n}(x)\right)\widehat{U}^\dagger(x) + \frac{1}{2}\sum_{x \neq x'}\sum_{nm} \widehat{\mathcal{J}}_n(x)\widehat{\mathcal{T}}_{nm}\widehat{\mathcal{J}}_m^\dagger(x') + h.c. \quad [3.1]$$

Here we have generalized the coupling term to allow more complicated couplings between the local site-specific operators and

$$\widehat{\mathcal{J}}_n(x) = \widehat{U}(x)\widehat{\mathcal{J}}_n\widehat{U}^\dagger(x) \quad [3.2]$$

The effective single-site Hamiltonian in SCEHT is then:

$$\widehat{\mathcal{H}}_{eff}(x) = \widehat{U}(x)\left(\widehat{H}(x) - \mu\hat{n}(x)\right)\widehat{U}^\dagger(x) + \sum_n \widehat{\mathcal{J}}_n(x) \cdot \widehat{\mathcal{B}}_n(x) + h.c. \quad [3.3]$$

$$\widehat{\mathcal{B}}_n(x) = \sum_{m, x' \neq x} \widehat{\mathcal{T}}_{nm}\langle\widehat{\mathcal{J}}_m^\dagger(x')\rangle \quad [3.4]$$

Here the expectation is taken over the self-consistent ground state.

With Eqs.[3.1]-[3.4], we then have the following generalized SCEHT for the ground state that goes beyond the gauge fixing condition of the original SCEHT of Eq.[1.6]:

$$\mathcal{E}_G = argmin_{\widehat{U}(x)} \sum_x \left(\langle\widehat{\mathcal{H}}_{eff}(x)\rangle - \frac{1}{2}\left(\sum_n \langle\widehat{\mathcal{J}}_n(x)\rangle \cdot \widehat{\mathcal{B}}_n(x) + h.c.\right)\right) \quad [3.5]$$

And again the expectation is taken over the self-consistent ground state of the effective Hamiltonian.

We propose to use Eq.[3.5]- Eq.[3.5] and apply the following numerical procedure to find the true ground state energy and gauge condition that goes beyond the original SCEHT (Eq.[1.6]).

1. For a cluster of $x$ in a cube, randomly generate unitary matrix $\widehat{U}(x)$, which can be specified by a set of parameters $\{\alpha_i(x)\}$. The periodic boundary condition is imposed for $\widehat{U}(x)$.
2. For the fixed distribution of $\widehat{U}(x)$ (constrain), solve for the self-consistent effective Hamiltonian problem [3.3] and [3.4] iteratively.
3. Calculate the self-consistent ground state energy of [3.5].
4. Record the configuration and the associated constrained ground state energy $\varepsilon\{\alpha_i(x)\}$.

5. Randomly choose another configuration, repeat steps 1-3, if the calculated total energy is lower than the previous constrained ground state energy on record, replace the record.
6. Repeat 5 until the lowest constrained ground state energy converges.

Finally in the appendix, we present the results for non-interacting fermion ground state using the SCEHT.

## Conclusion

We have presented a rigorous mathematical foundation to the SCEHT and in a more general manner, reinterpreted gauge field theory based on two Ansatz, namely the Ansatz of Locality and Ansatz of Continuity. Gauge fields are introduced and correct total energy functional in relations to the coupling of gauge field is given. We also provides a Monte-Carlo numerical scheme for the search of the ground state that goes beyond the SCEHT.

## Appendix

In this appendix, we would like to present the result of local density matrix for non-interacting fermi sea, for example, for a single band Hubbard Model with onsite $U = 0$.

Following SCEHT, we have ($t > 0$):

$$H_{eff}(0) = \begin{pmatrix} 0 & -t \cdot \left(\sum_n \langle \psi_n \rangle^*\right) \\ -t \cdot \left(\sum_n \langle \psi_n \rangle\right) & -\mu \end{pmatrix} = \begin{pmatrix} 0 & -\tilde{t}^* \\ -\tilde{t} & -\mu \end{pmatrix}$$

The ground state of is then

$$\begin{pmatrix} -\dfrac{\tilde{t}^*}{\sqrt{\varepsilon_g^2 + |\tilde{t}|^2}} \\ \dfrac{\varepsilon_g}{\sqrt{\varepsilon_g^2 + |\tilde{t}|^2}} \end{pmatrix}, \varepsilon_g = \frac{-\mu - \sqrt{\mu^2 + 4|\tilde{t}|^2}}{2}$$

The self-consistent condition for the SCEHT is then:

$$\langle \psi \rangle = -\frac{\varepsilon_g t \sum_n \langle \psi_n \rangle}{\varepsilon_g^2 + t^2 |(\sum_n \langle \psi_n \rangle)|^2}, \quad \langle n \rangle = \frac{\varepsilon_g^2}{\varepsilon_g^2 + t^2 |(\sum_n \langle \psi_n \rangle)|^2}$$

For non-interacting system, it is natural to assume translational symmetry, thus $\langle \psi_n \rangle = \phi e^{i\theta_n}$,

The self-consistent condition then gives

$$\phi^2 = \frac{-\varepsilon_g t e^{-i\theta_0} \sum_n e^{i\theta_n} - \varepsilon_g^2}{t^2 |\sum_n e^{i\theta_n}|^2}, \qquad \langle n \rangle = \frac{\varepsilon_g^2}{\varepsilon_g^2 + t^2 \phi^2 |(\sum_n e^{i\theta_n})|^2} \qquad [A.1]$$

$$\varepsilon_g = \frac{-\mu - \sqrt{\mu^2 + 4t^2 \cdot \phi^2 |(\sum_n e^{i\theta_n})|^2}}{2} \qquad [A.2]$$

Define geometrical:

$$\zeta = e^{-i\theta_0} \sum_n e^{i\theta_n}$$

A real number, the solution is

$$\varepsilon_g = -\frac{\mu + t\phi\zeta}{2} \qquad [A.3]$$

The total energy per site is thus:

$$\mathcal{E}_G = \varepsilon_g - \frac{\varepsilon_g t^2 \phi^2 \zeta^2}{\varepsilon_g^2 + t^2 \phi^2 \zeta^2} = \frac{\varepsilon_g^3}{\varepsilon_g^2 + t^2 \phi^2 \zeta^2} = -\frac{\varepsilon_g^2}{t\phi\zeta} = -\frac{(\mu + t\phi\zeta)^2}{4t\phi\zeta} \qquad [A.3]$$

And $\zeta$ is solved by $\langle n \rangle = \frac{\varepsilon_g^2}{\varepsilon_g^2 + t^2 \phi^2 \zeta^2} = -\frac{\varepsilon_g}{t\phi\zeta} = \frac{\mu + t\phi\zeta}{2t\phi\zeta}$ to be

$$\phi\zeta = \frac{\mu}{t(2\langle n \rangle - 1)}$$

Substitute this into [A.3] we arrive at

$$\mathcal{E}_G = -\frac{\mu}{2} \cdot \frac{\langle n \rangle}{2\langle n \rangle - 1} = -\frac{\langle n \rangle}{2} \zeta \phi(\zeta) t \qquad [A.4]$$

The lowest ground state energy is thus attained when $\phi(\zeta)\zeta$ is the highest for each value of $\zeta$

Given $\phi\zeta$ the chemical potential and occupancy number is:

$$\mu = t(2\langle n \rangle - 1)\phi\zeta$$

So when $\langle n \rangle < \frac{1}{2}, \mu < 0$; $\langle n \rangle > \frac{1}{2}, \mu > 0$.

Finally $z(\zeta) = \phi(\zeta)\zeta$ is solved as the positive root of the following quadratic equation by substituting [A.3] into [A.1]:

$$z^2 = \frac{(\mu + tz)(2t\zeta - \mu - tz)}{4t^2}$$